\documentclass[12pt, titlepage, reqno]{article}

\usepackage{graphicx}
\usepackage{amsmath,amsfonts, latexsym}
\usepackage{mathrsfs}
\usepackage{mathabx}
\usepackage{bm}
\usepackage{mathrsfs}
\usepackage{bbm}
\usepackage{subfigure}
\usepackage{authblk}
\usepackage[super,comma]{natbib}
\usepackage{indentfirst}
 \usepackage[mathscr]{eucal}
 \usepackage{enumerate}
 \usepackage{lineno,xcolor}
 \usepackage{fancyhdr}

\makeatletter

\newcommand{\Rmnum}[1]{\expandafter\@slowromancap\romannumeral #1@}
\newcommand{\onlinecite}[1]{\hspace{-1 ex} \nocite{#1}\citenum{#1}}

\renewcommand\@biblabel[1]{\textsuperscript{#1}}
\makeatother

\begin{document}

\begin{titlepage}

\begin{center}
\thispagestyle{fancy}
\textbf{Fano resonance scattering in waveguide with an impedance boundary condition}

\vspace{10ex}

Lei Xiong\footnote{lei.xiong@univ-lemans.fr}, Wenping Bi, and Yves Aur\'{e}gan\\

Laboratoire d'Acoustique de l'Universit\'e du Maine\\

UMR CNRS 6613, \\

Avenue Olivier  Messiaen, 72085 Le Mans Cedex 9, France

\today
\end{center}

\end{titlepage}


\begin{abstract}
\addtocounter{page}{1}
\thispagestyle{fancy}

Sound propagation in a waveguide lined with one section of locally reactive material is studied by resonance scattering approach. The objective is to understand the effects of mode coupling in the lined section on the transmission. It is  shown that a transmission zero is present in the vicinity of a resonance peak when a numerically real resonance frequency of the \textit{open} lined section (opened to infinities through the rigid parts of the waveguide) is crossed. The transmission zero and immediate resonance peak form a Fano resonance, it has been explained as an interaction between a resonance and the non-resonant background. The real resonance frequency  and its corresponding trapped mode are formed by the interferences (couplings) between two neighbor modes with complex resonance frequencies. It is also linked to the avoided crossing of eigenvalues and the exceptional point. The scattering matrix is expressed in terms of a matrix $\mathsf{H_{eff}}$ which describes approximately the complex resonances in the \textit{open} lined section. With the aid of the eigenvalues and eigenfunctions of matrix $\mathsf{H_{eff}}$,  the traditional acoustic resonance scattering formula can be extended to describe the coupling effects between the \textit{open} lined section and the rigid parts of the waveguide.
\\
\\


\end{abstract}

\addtocounter{page}{2}

\section{\label{sec1:level1}INTRODUCTION}
 
\setlength{\parindent}{5ex}
 
Acoustic liners are commonly used in noise control devices for duct systems. Typical applications include silencers for ventilation systems, wall treatments for aircraft engines, and silencers for industrial gas turbines. To increase the liner efficiency, various strategies can be used. One is to find a new material design for which the impedance is close to the optimal value in the targeted range of frequencies\cite{beck2013impedance, groby2015use}. Another strategy is to take advantage of the acoustic impedance changes (like discontinuities) in axial\cite{lansing, unruh, koch1} or circumferential\cite{fuller, watson} segments, or both of them\cite{regan, bi_modelling_2006}. The strategy used in this paper is different: The idea is to couple the incoming propagative mode in the waveguide with the modes localized in the lined region. Those localized modes exist for particular values of the geometry in a uniform liner but can also be created by impedance variations. As an example, a very simple  2D model with an uniform liner is analyzed in this paper.

 \medskip

Such interferences between resonances and scattering appear in Fano resonances (for review articles see, for example, Refs.\;\onlinecite{lukyanchuk_fano_2010}\;and\;\onlinecite{ miroshnichenko_fano_2010}, and the references therein). In contrast to the conventional isolated resonances scattering, Fano resonance is explained by constructive and destructive interferences between a resonance (trapped mode) or a complex resonance (quasi-trapped mode) and the background or nonresonant scattering\cite{fano_effects_1961}. A transmission zero is produced as a real resonance frequency is crossed\cite{hein_trapped_2012}.

 \medskip
 
One of the crucial ingredients to form a Fano resonance is to have trapped mode with a real resonance frequency. Trapped modes are localized oscillations in unbounded media and do not radiate energy to infinity.  They were first observed experimentally in acoustics by Parker in 1966\cite{parker_resonance_1966}. Discrete trapped modes may exist below the first cut-off frequency of the transverse modes, provided that  some kinds of defect or variations of geometry exist\cite{duan_complex_2007, pagneux_trap}. Discrete trapped modes may also exist above the first cut-off frequency for specific parameter combinations, they are called embedded trapped modes \cite{duan_complex_2007} or bound states in continuum (BIC) in quantum mechanics\cite{neumann1}.  Friedrich and Wintgen\cite{friedrich_interfering_1985} demonstrated that BIC is a natural feature of common physical situations, and can occur due to the interference of resonances. They have linked BICs directly to the phenomenon of an avoided crossing of neighbored resonance states (modes with complex eigenvalues). Recently, BICs in the vicinity of exceptional points were also found in open quantum billiards\cite{sadreev_bound_2006}. Sadreev $et\ al.$\cite{sadreev_bound_2006} found that the BICs are close to the points of degeneracy of the closed quantum system. When the system is opened, the degeneracies are lifted and avoided crossings occur.

 \medskip
 
Fano resonance scatterings in acoustics and their relations with trapped modes in waveguides including obstacles or in duct-cavity systems have been studied by Hein $et\ al.$\cite{hein_trap, hein_trapped_2012}. They used finite-element method to compute numerically the acoustic resonances as well as transmission and reflection for an incoming duct mode. They obtained complex resonance frequencies and corresponding eigenfunctions, the homogeneous solutions of the Helmholtz equation with absorbing boundary conditions. The complex resonance frequencies are the positions of the poles of scattering matrix of the corresponding scattering problem. Fano resonance scatterings were related to three types of trapped modes: antisymmetric (about duct axis) trapped modes below the first cut-off frequency, embedded trapped modes linked to avoided crossings of resonances, and trapped modes associated with Fabry-P\'erot interferences between cavities. 

 \medskip
 
In this paper, we study sound propagation in a waveguide lined on a section with a locally reacting material by the resonance scattering approach\cite{uberall}. The objective is to understand the effects of mode coupling on the transmission of the lined section. We show that by varying a control parameter (the section length or the product of the lined admittance and the frequency), two neighbored modes with complex resonance frequencies interfere in the scattering region: the lined section opened to the two semi-infinite waveguides. In the vicinity of an exceptional point, where the eigenvalues and eigenfunctions coalesce, one mode turns to be trapped, the corresponding resonance frequency (eigenvalue) is real.  A transmission zero is present in the vicinity of a resonance peak when the real resonance frequency is crossed, this is also called Fano resonance  (section \Rmnum{3}).

 \medskip
 
In section \Rmnum{2}, we derive the Scattering matrix ($\mathsf{S}$ matrix). For that, we project the Helmholtz equation over the eigenfunctions of the rigid closed counterpart of the \textit{open} lined section which form an orthogonal and complete function basis. The used eigenfunctions include the transverse and axial components, thus this method generalizes the Multimodal method\cite{bi_modelling_2006} in which the wave function is expanded only in terms of transverse eigenfunctions. We express the scattering matrix $\mathsf{S}$ in terms of an effective matrix $\mathsf{H_{eff}}$. The matrix $\mathsf{H_{eff}}$ approximately describes the complex resonances of the \textit{open} lined section. Its eigenvalues are complex and give the poles of the $\mathsf{S}$ matrix. With the help of its eigenfunctions, we can extend the traditional acoustic resonance scattering formula\cite{uberall} to describe the coupling effects between the \textit{open} lined section and the rigid waveguide.

\section{\label{sec2:level1} MODEL}

\begin{figure}[htb]
\centering
\subfigure{}{\label{conf_2}
\includegraphics[height=5cm,width=8cm]{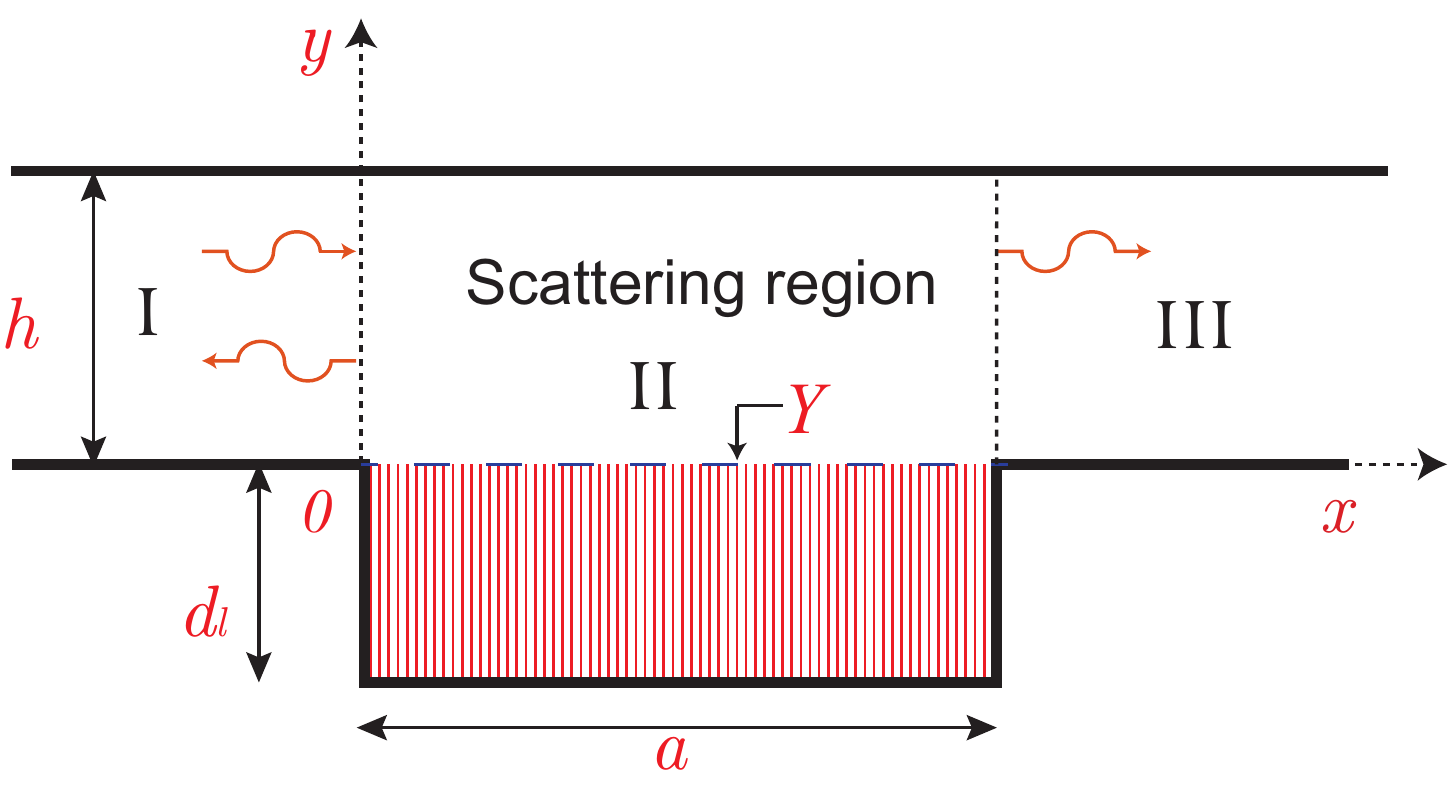}}
\caption{(Color online) A two-dimensional infinite waveguide lined with a locally reacting  material of length $a$. $Y$ refers to the admittance. The closed counterpart of the scattering region is called \textit{closed cavity}, with rigid conditions at $x=0$ and $x=a$.}
\label{Figure1}
\end{figure}

We consider the acoustic scattering problem in a two-dimensional  infinite waveguide lined over a finite length $a$ with a locally reacting material, as shown in Fig.\;\ref{Figure1}. The waveguide is decomposed into three parts: two semi-infinite rigid waveguides $x\leq 0$ and $x\geq a$ (regions  \Rmnum{1} and  \Rmnum{3}, respectively),  and one scattering region (region  \Rmnum{2}). Time dependence is assumed as exp$(-\text{j}\omega t)$ and will be omitted in the following. 
In the following, all the quantities are non-dimensionalized. The sound pressure $p(x,y)$ in the waveguide satisfies the non-dimensional Helmholtz equation
\begin{equation}
\left(\frac{\partial^2}{\partial x^2}+\frac{\partial^2}{\partial y^2}+K^2\right)p(x,y)=0,
\label{goven}
\end{equation}
where $K=(\omega/c_0)h$ refers to dimensionless frequency, $\omega$ is the frequency, $c_0$ is the sound velocity, and $h$ is the height of the waveguide.  Sound pressure $p$ and coordinates ($x$ and $y$) are normalized by $\rho_0{c_0}^2$ and $h$, respectively, with $\rho_0$ the air density. The transverse  boundary conditions in regions \Rmnum{1}, \Rmnum{3}, and  \Rmnum{2} are 
\begin{equation}
\frac{\partial p}{\partial y}\Bigm\vert_{y=0\; \text{and}\; 1}=0 \; \text{(for regions \Rmnum{1} and \Rmnum{3})} \;\;\;\;
\end{equation}
and
\begin{equation}
 \frac{\partial p}{\partial y}\Bigm\vert_{y=0}=-\text{j}KYp\; \text{and}\; \frac{\partial p}{\partial y}\Bigm\vert_{y=1}=0 \; \text{(for region \Rmnum{2})},
 \label{BC_III}
\end{equation}
respectively.

 \medskip
 
Inspired by the R-matrix method\cite{racec_evanescent_2009,racec_fano_2010},  the sound pressure $p(x,y)$ in region \Rmnum{2} is expanded in terms of an orthogonal and complete set of functions $\psi_{\mu\nu}(x,y)$
\begin{equation}
p(x,y)=\sum_{\mu=0}^{N_x-1}\sum_{\nu=0}^{N_y-1}a_{\mu\nu}\psi_{\mu\nu}(x,y)=\bm{\psi}^T\bm{a}, \label{p_in}
\end{equation}
where the sums have been truncated by $N_x$ and $N_y$, ``$^T$" refers to transpose. $\bm{\psi}$ is a column vector, its elements are arranged as $(\mu, \nu)=(0, 0), (0, 1), ..., (0, N_y-1), (1, 0), (1, 1), ..., (1, N_y-1), ..., (N_x-1, 0), (N_x-1, 1), ..., (N_x-1, N_y-1)$. We choose $\psi_{\mu,\nu}$ to be the eigenfunctions of closed \textit{rigid} cavity defined by
\begin{eqnarray}
 &\nabla^2\psi_{\mu\nu}=-\gamma^2_{\mu\nu}\psi_{\mu\nu},& \label{rigid cavity}\\
 &\frac{\partial \psi_{\mu\nu}}{\partial x}\Bigm\vert_{x=0\; \text{and}\; a/h}=0, \;\; \frac{\partial \psi_{\mu\nu}}{\partial y}\Bigm\vert_{y=0\;\text{and}\; 1}=0. \label{rigid cavity bound}
\end{eqnarray}
By solving the eigenproblem of Eqs.\;(\ref{rigid cavity}) and (\ref{rigid cavity bound}), the eigenfunctions $\psi_{\mu\nu}$ and eigenvalues $\gamma_{\mu\nu}$ are given as $\psi_{\mu\nu}=(1/\sqrt{\Lambda_{x}\Lambda_{y}})\cos(\mu\pi x/(a/h))\cos(\nu\pi y)$ and $\gamma_{\mu\nu}=\sqrt{(\mu\pi)^2/(a/h)^2+(\nu\pi)^2},$ respectively, where $\Lambda_x$ and $\Lambda_y$ are normalization coefficients.

 \medskip
 
Multiplying Eq.\;(\ref{goven}) by $\bm{\psi}$, integrating over $x$ and $y$, applying Green's theorem, and using Eq.\;(\ref{rigid cavity}) and the boundary conditions Eq.\;(\ref{rigid cavity bound}), we obtain
\begin{equation}\label{bound-general}
p(x, y)=\int_0^1 \left[R(x, y, 0, y')\frac{\partial p(x', y')}{\partial x'}\Bigm\vert_{x'=0}-R(x, y, a, y')\frac{\partial p(x', y')}{\partial x'}\Bigm\vert_{x'=a/h}\right]dy',
\end{equation}
(see Appendix), where
\begin{equation}\label{R-function}
R(x, y, x', y') =\bm{\psi}^T(x, y)\left[{K^2\mathsf{I}-\mathsf{H_{in}}}\right]^{-1}\bm{\psi}(x', y').
\end{equation}
In Eq.\;(\ref{R-function}), $\mathsf{I}$ refers to identity matrix, $\mathsf{H_{in}}=\mathsf{\Gamma}-\text{j}KY\mathsf{C^{in}}$, where  $\mathsf{\Gamma}$ is a diagonal matrix with elements $\gamma_{\mu\nu}^2$,   and  $\mathsf{C^{in}}$ is a block diagonal matrix, its elements  can be calculated analytically by
\begin{equation}
C^{in}_{\mu\nu, \mu'\nu'}=\int_0^{a/h}\psi_{\mu\nu}(x', 0)\psi_{\mu'\nu'}(x', 0)dx'.
\end{equation} 
With the help of Eq.\;(\ref{Eq_A5}), it can be noted that the eigenfunctions of the matrix  $\mathsf{H_{in}}$ are the modes of the closed cavity with the admittance on the bottom wall.

 \medskip
Equation (\ref{bound-general}) links the sound pressures in the scattering region (\Rmnum{2}) with their first derivatives with respect to $x$ at the interfaces between the scattering region and the other two regions (\Rmnum{1} and \Rmnum{3}).

 \medskip
 
The sound pressure is written as a sum of the incident (amplitudes $c_m$) and reflected modes for region \Rmnum{1}, and only transmitted modes for region \Rmnum{3} 
 \begin{equation}
p(x,y)=\left\{\begin{aligned}&\sum_{m=0}^{M-1}c_me^{\text{j}K^x_{m}x}\phi_{m}(y)+\sum_{m=0}^{M-1}\sum_{m'=0}^{M-1}\mathsf R_{m,m'}c_me^{-\text{j}K^x_{m'}x}\phi_{m'}(y), &x\leq 0,\\
&\sum_{m=0}^{M-1}\sum_{m'=0}^{M-1}\mathsf T_{m,m'}c_me^{\text{j}K^x_{m'}(x-a/h)}\phi_{m'}(y), & x\geq a/h, \label{p1} \end{aligned} \right. 
\end{equation}
where $\mathsf{R}_{m,m'}$ and $\mathsf{T}_{m,m'}$ refer to the reflection and transmission coefficients, and $M$ is the truncation number. $K^x_{m}=\sqrt{K^2-\alpha_{m}^2}$ are axial wavenumbers of mode $m$ in the rigid ducts. $\alpha_{m}$ and $\phi_{m}$ are the eigenvalues and eigenfunctions of transverse modes in regions \Rmnum{1} and  \Rmnum{3}, they are given as $\alpha_{m}=m\pi$ and $\phi_{m}=\Lambda cos(\alpha_m y)$, with $\Lambda$ the normalization coefficients. Equation (\ref{p1}) is written in matrix form as
\begin{equation}\label{p1a}
p(x,y)=\left\{\begin{aligned}&\bm{\phi}^T\mathsf{E^+_0}\bm{c}+\bm{\phi}^T\mathsf{E^-_0}\mathsf{R}\bm{c}, &x\leq 0,\\
& \bm{\phi}^T\mathsf{E^+_a}\mathsf{T}\bm{c},& x\geq a/h,\end{aligned} \right. 
\end{equation}
where $\mathsf{E^+_0}$, $\mathsf{E^-_0}$, and $\mathsf{E^+_a}$ are ($M\times M$) diagonal matrices with the elements $e^{\text{j}K_m^xx}$, $e^{-\text{j}K_m^xx}$, and $e^{\text{j}K_m^x(x-a/h)}$, respectively. $\bm{\phi}$ is  a ($M\times 1$) column vector, its elements are the eigenfunctions $\phi_{m}$. 

 \medskip
 
Substituting Eq.\;(\ref{p1a}) into Eq.\;(\ref{bound-general}) at $x=0$ and $x=a/h$, we obtain
\begin{eqnarray}
\mathsf{I_M}+\mathsf{R} &=& \text{j}\mathsf{C_0}^T(K^2\mathsf{I}-\mathsf{H_{in}})^{-1}(\mathsf{C_0}\mathsf{K^x}(\mathsf{I_M}-\mathsf{R})-\mathsf{C_a}\mathsf{K^x}\mathsf{T}), \label{con1}\\
\mathsf{T} &=& \text{j}\mathsf{C_a}^T(K^2\mathsf{I}-\mathsf{H_{in}})^{-1}(\mathsf{C_0}\mathsf{K^x}(\mathsf{I_M}-\mathsf{R})-\mathsf{C_a}\mathsf{K^x}\mathsf{T}),\label{con2}
\end{eqnarray}
where $\mathsf{K^x}$ is a ($M\times M$) diagonal matrix with elements $K^x_m$, and $\mathsf{I_M}$ is a ($M\times M$) identity matrix.  The elements of matrices $\mathsf{C_0}$ and $\mathsf{C_a}$ are
\begin{equation}
C_{0(a), \mu\nu, m}=\int_0^1\psi_{\mu\nu}(x'=0(a/h), y')\phi_m(y')dy',
\end{equation}
where $\mathsf{C_0}$ and $\mathsf {C_a}$ have dimensions $N\times M$ with $N=N_x\cdot N_y$.  From Eqs. (\ref{con1}) and (\ref{con2}), we obtain
\begin{equation}\label{rt}
\left[\begin{array}{c}\mathsf{R}\\ \mathsf{T}\end{array}\right]=\left[{\mathsf{I_{2M}}+\text{j}\mathsf{G}\mathsf{K^x_{2M}}}\right]^{-1}\left\{-[\mathsf{I_M}, \mathsf{0_M}]^T+\text{j}\mathsf{C_{0a}}^T[K^2\mathsf{I}-\mathsf{H_{in}}]^{-1}\mathsf{C_0}\mathsf{K^x}\right\},
\end{equation}
where $\mathsf{G}=\mathsf{C_{0a}}^T(K^2\mathsf{I}-\mathsf{H_{in}})^{-1}\mathsf{C_{0a}}$, $\mathsf{C_{0a}}=[\mathsf{C_0}, \mathsf{C_a}]$ is a ($N\times 2M$) matrix, $\mathsf{I_{2M}}$ is a ($2M\times 2M$) identity matrix, $\mathsf{K^x_{2M}}$ is a ($2M\times 2M$) block diagonal matrix with two ($M\times M$) diagonal matrices $\mathsf {K^x}$ on its main diagonal, and $\mathsf{0_M}$ is a ($M\times M$) zero matrix.

 \medskip
 
If we assume that the problem is symmetric, the scattering matrix can be written as
\begin{equation}
\mathsf{S}=\left[\begin{array}{cc}\mathsf{R} & \mathsf{T}\\ \mathsf{T} & \mathsf{R}\end{array}\right]=\left[\mathsf{I_{2M}}+\text{j}\mathsf{G}\mathsf{K^x_{2M}}\right]^{-1}\left[{-\mathsf{I_{2M}}+\text{j}\mathsf{G}\mathsf{K^x_{2M}}}\right].
\label{Eq_20}
\end{equation}
The  scattering matrix $\mathsf{S}$ can be expressed as\cite{stockmann_effective_2002} 
\begin{equation}
\mathsf{S}=-\mathsf{I_{2M}}+2\text{j}\mathsf{C_{0a}}^T\left[K^2\mathsf{I}-\mathsf{H_{eff}}\right]^{-1}\mathsf{C_{0a}}\mathsf{K^x_{2M}},
\label{S4}
\end{equation}
(see Appendix), where 
\begin{equation}
\mathsf{H_{eff}}=\mathsf{H_{in}}-\text{j}\mathsf{C_{0a}K^x_{2M}}\mathsf{C_{0a}}^T\label{Heff}.
\end{equation}
The eigenvalues $K_\lambda$ and eigenfunctions $\tilde{\varphi}_\lambda=\bm{\psi}^T\bm{V}_\lambda$ are defined by the eigenproblem of matrix $\mathsf{H_{eff}}$, 
\begin{equation}
\mathsf{H_{eff}}\bm{V}_\lambda=K^2_\lambda\bm{V}_\lambda.
\label{eig_heff}
\end{equation}
They describe the complex resonances of scattering region \Rmnum{2}, which is opened to infinities through regions \Rmnum{1} and \Rmnum{3}, and truncated at the interfaces, $x=0$ and $x=a/h$. The elements of vector  $\bm{\psi}$ are the eigenfunctions of the $rigid$ closed cavity defined by Eqs.\;(\ref{rigid cavity}) and (\ref{rigid cavity bound}). Because the eigenfunctions $\tilde{\varphi}_\lambda$ in scattering region \Rmnum{2} are non-separable in $x$ and $y$, we use only one index $\lambda$ to describe the eigenvalues $K_\lambda$ and the eigenfunctions $\tilde{\varphi}_\lambda$. It is noted that the complex resonances are calculated in Refs.\;\onlinecite{hein_trap} and\;\onlinecite{ hein_trapped_2012} by finite element method with absorbing boundary conditions.

 \medskip
 
With the help of eigenvectors  $\bm{V}_\lambda$, Eq.\;(\ref{S4}) can be written as
\begin{eqnarray}\label{Sa}
\mathsf{S} &=& -\mathsf{I_{2M}}+2\text{j}\mathsf{C_{0a}}^T\mathsf{V}[K^2\mathsf{V}\mathsf{V}^{-1}-\mathsf{V}\mathsf{H_{eff}}\mathsf{V}^{-1}]^{-1}\mathsf{V}^T\mathsf{C_{0a}}\mathsf{K^x_{2M}}\\\nonumber
&=& -\mathsf{I_{2M}}+2\text{j}\mathsf{\tilde{C}_{0a}}^T[K^2\mathsf{I}-\mathsf{K_\lambda}]^{-1}\mathsf{\tilde{C}_{0a}}\mathsf{K^x_{2M}},
\end{eqnarray}
where $\mathsf{K_\lambda}$ is a ($N\times N$) diagonal matrix with $K_\lambda^2$ its main elements. $\mathsf{V}$ is a ($N\times N$)  matrix with its columns the eigenvectors $\bm{V}_\lambda$ of matrix  $\mathsf{H_{eff}}$. $\mathsf{H_{eff}}$ is a symmetric non-Hermitian matrix, its eigenvectors are bi-orthogonal. Hence, $\mathsf{V}^{-1}=\mathsf{V}^T$ has been used in Eq.\;(\ref{Sa}). Matrix $\mathsf{\tilde{C}_{0a}}$ is defined as
\begin{eqnarray}
\mathsf{\tilde{C}_{0a}}=\mathsf{V}^T\mathsf{C_{0a}} &=& \int_0^h\mathsf{V}^T[\bm{\psi}(x'=0, y')\bm{\phi}^T(y'), \bm{\psi}(x'=a/h, y')\bm{\phi}^T(y')]dy'\\\nonumber
&=& \int_0^h[\bm{\tilde{\varphi}}(x'=0, y')\bm{\phi}^T(y'), \bm{\tilde{\varphi}}(x'=a/h, y')\bm{\phi}^T(y')]dy',
\end{eqnarray}
where $\bm{\tilde{\varphi}}(x'=0, y')$ and $\bm{\tilde{\varphi}}(x'=a/h, y')$ are $(N\times 1)$ vectors with elements  the eigenfunctions $\tilde{\varphi}_\lambda$ of scattering region \Rmnum{2} at $x'=0$ and $x'=a/h$, respectively. Matrix $\mathsf{\tilde{C}_{0a}}$ describes the couplings of the scattering region \Rmnum{2} with regions \Rmnum{1} and \Rmnum{3}.  Equation\;(\ref{Sa}) is not valid at exceptional points at which  the eigenvalues and eigenfunctions coalesce, and therefore $\mathsf{V}^{-1}$ is singular.  

 \medskip
 
Equation (\ref{Sa}) extends the traditional acoustic resonance scattering formula\cite{uberall}, in which the complex eigenfrequencies of scattering region give the resonance poles, to describe the coupling effects between scattering region and rigid waveguides. 

\section{\label{sec3:level1}RESULTS AND DISCUSSIONS}

First, we show how the trapped modes with real resonance frequencies occur in the vicinity of exceptional points and how they are linked to avoided crossings of the eigenvalues of matrix $\mathsf{H_{eff}}$. Second, we  show that a transmission zero is present in the vicinity of a resonance peak when the real resonance frequency is crossed. Finally, we consider the effects of dissipation in acoustic absorbing material. The liner can be described by an impedance model ($Y=1/Z$):  
\begin{equation}\label{impedancemodel}
Z=Re+\text{j}\cot(Kd_l/h),
\end{equation}
where $d_l/h$ is the normalized depth of the liner, and $Re$ is the resistance. $Z$ is assumed to be uniform. All the numerical results are calculated with the truncation number $M=30, N_x=30$, and $N_y=30$. Although Eqs.\;(\ref{S4}) and (\ref{Sa}) are valid for arbitrary multimodal incidences, for the sake of simplicity, we assume that only plane waves are  incident in the following.

 \medskip

 \begin{figure}[h!]
\centering
\subfigure{}{
\includegraphics[height=6.5cm,width=8.3cm]{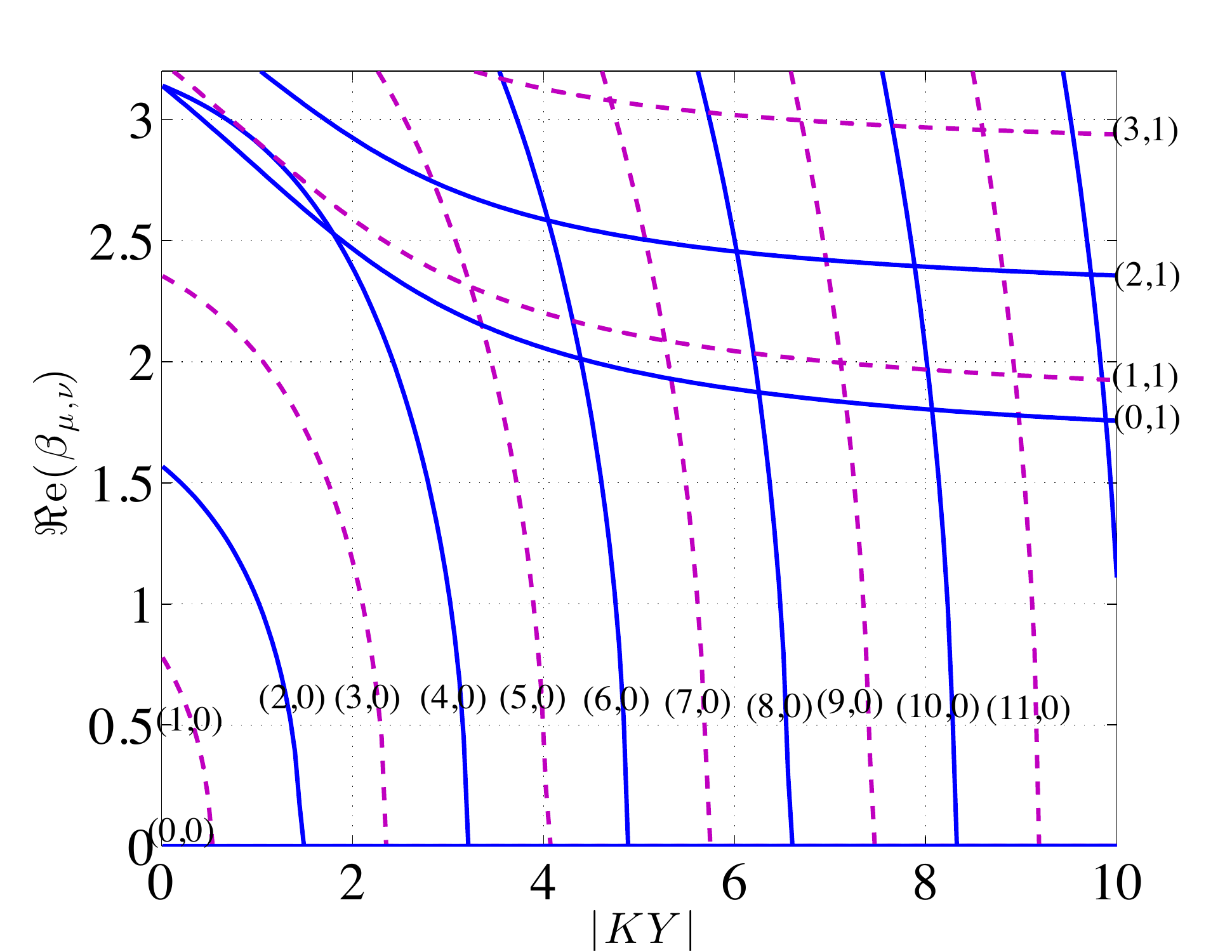}}
\caption{(Color online) Trajectories of the real parts of eigenvalues $\mathsf{\beta}_{\mu,\nu}$ of the closed cavity with admittance as  a function of $\vert KY\vert$ with $Re=0$ and $a/h=4$. Solid lines label even modes in $x$-direction, dashed lines label odd modes in $x$-direction. The mode indices ($\mu$, $\nu$) are marked near each curve.}
\label{Figure2}
\end{figure}
The eigenvalues of matrix $\mathsf{H_{in}}$, denoted $\beta_{\mu,\nu}^2$, represent the eigenvalues  of the closed cavity with the admittance. The trajectories of $\Re$e$(\beta_{\mu,\nu})$  as a function of $\vert KY\vert$ are shown in Fig.\;\ref{Figure2} with $a/h=4$ and $Re=0$. When the eigenvalue curves cross,  the modes are degenerate\cite{morse}. At the degeneracies, eigenvalues coalesce, while eigenfunctions are still bi-orthogonal\cite{morse} and there is no interaction between the modes.

\begin{figure}[h!]
\centering
\subfigure{}{
\includegraphics[height=6.5cm,width=8.3cm]{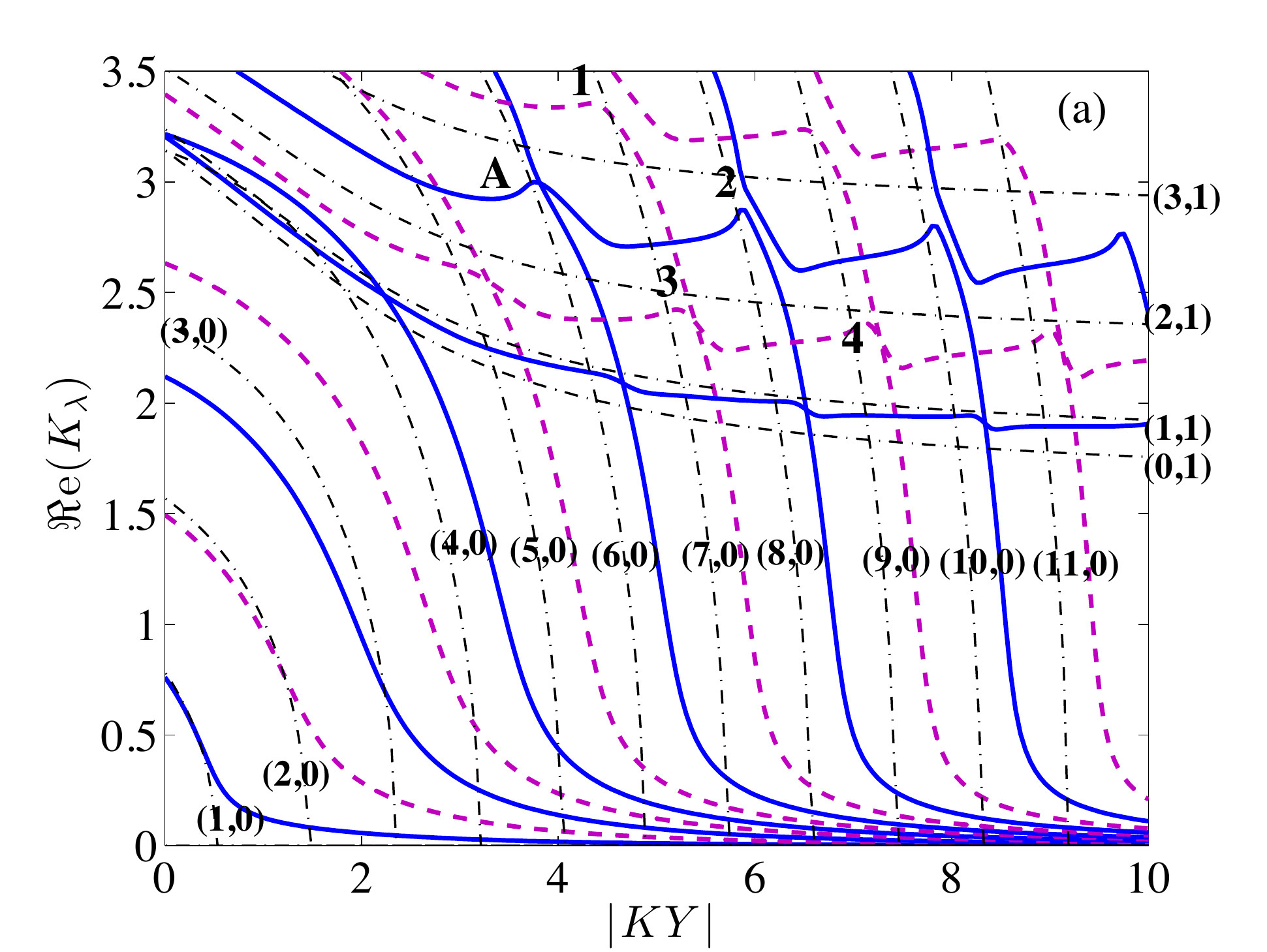}}
\subfigure{}{
\includegraphics[height=5cm,width=8.3cm]{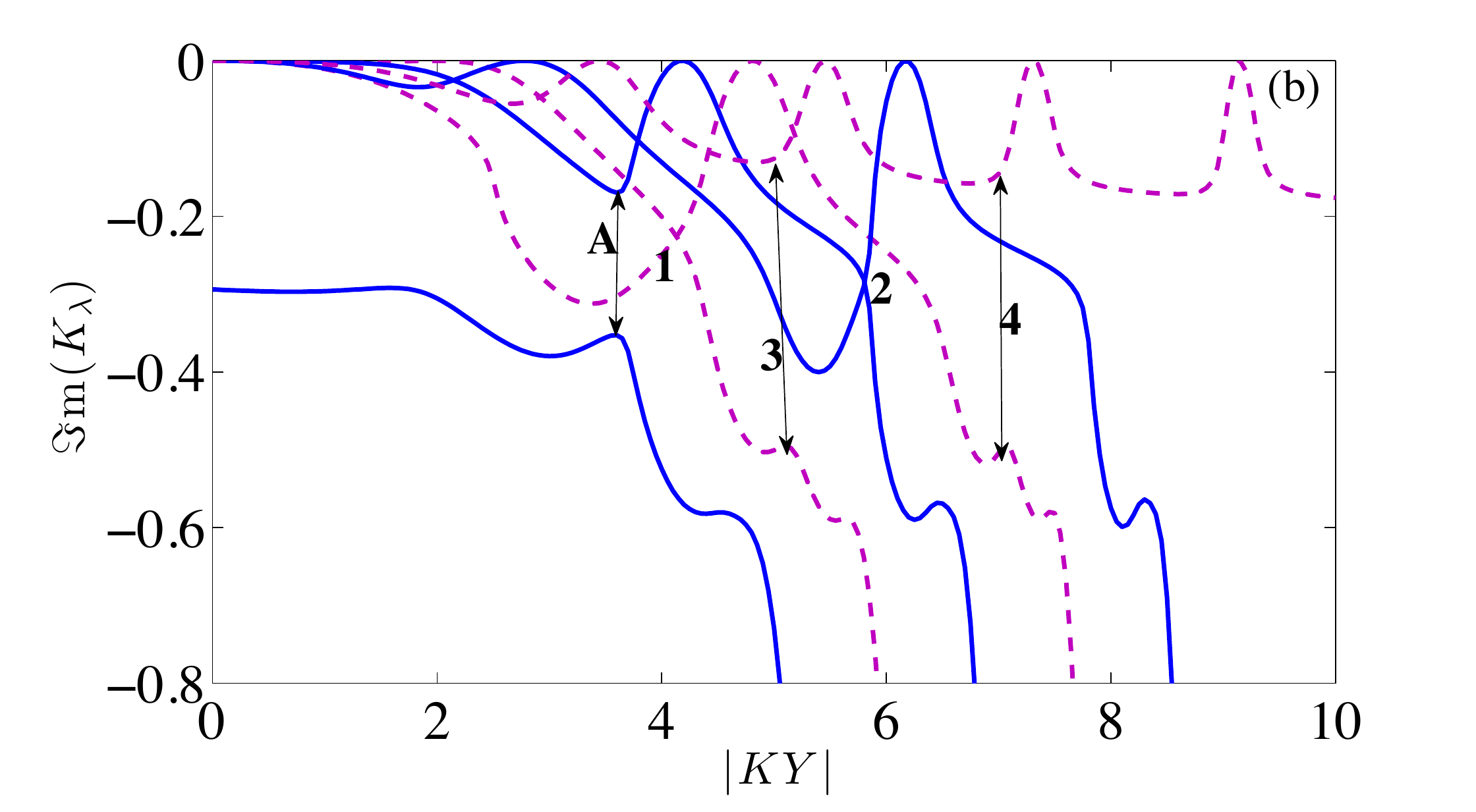}}
\caption{(Color online) Trajectories of $\Re $e$(K_\lambda)$ and $\Im $m$(K_\lambda)$ of matrix $\mathsf{H_{eff}}$ as a function of $\vert KY\vert$ with $K=2.5$, $a/h=4$,  and $Re=0$: (a) real parts and  (b) imaginary parts. Solid lines and dashed lines correspond to different symmetries as shown in Fig.\;\ref{Figure2}. For comparison, we plot also the trajectories of $\Re $e$(\beta_{\mu,\nu})$ of the closed cavity by dash-dot lines.}
\label{Figure3}
\end{figure}

We consider the modes behaviour when the region \Rmnum{2} is opened.  It can be modelled by matrix $\mathsf{H_{eff}}$ which depends on three parameters: $a/h$, $KY$, and $K$. This is due to the $K$ dependences of both the impedance boundary condition $KY$ and the effective radiation condition at the interfaces $x=0$ and $x=a/h$ (the term $\mathsf{K^x_{2M}}$ in Eq.\;(\ref{Heff})).  In Fig.\;\ref{Figure3}, we plot the  eigenvalue trajectories as a function of $\vert KY\vert$ with $a/h=4$ and $K=2.5$.  The curves with the same symmetry about $x$, which had crossings before in Fig.\;\ref{Figure2}, now have avoided crossings in Fig.\;\ref{Figure3}. The avoided crossings  of the eigenvalues $K_\lambda$ in Fig.\;\ref{Figure3} occur in the vicinity of the degeneracies of the closed cavity, for example those labelled by A, 1, 2, 3, and 4, respectively. Avoided crossings, already known in structural dynamics\cite{leissa, kuttler, perkins}, are less used in cavities lined with impedance.   

\begin{figure}[h!]
\centering
\includegraphics[height=7.5cm,width=8.3cm]{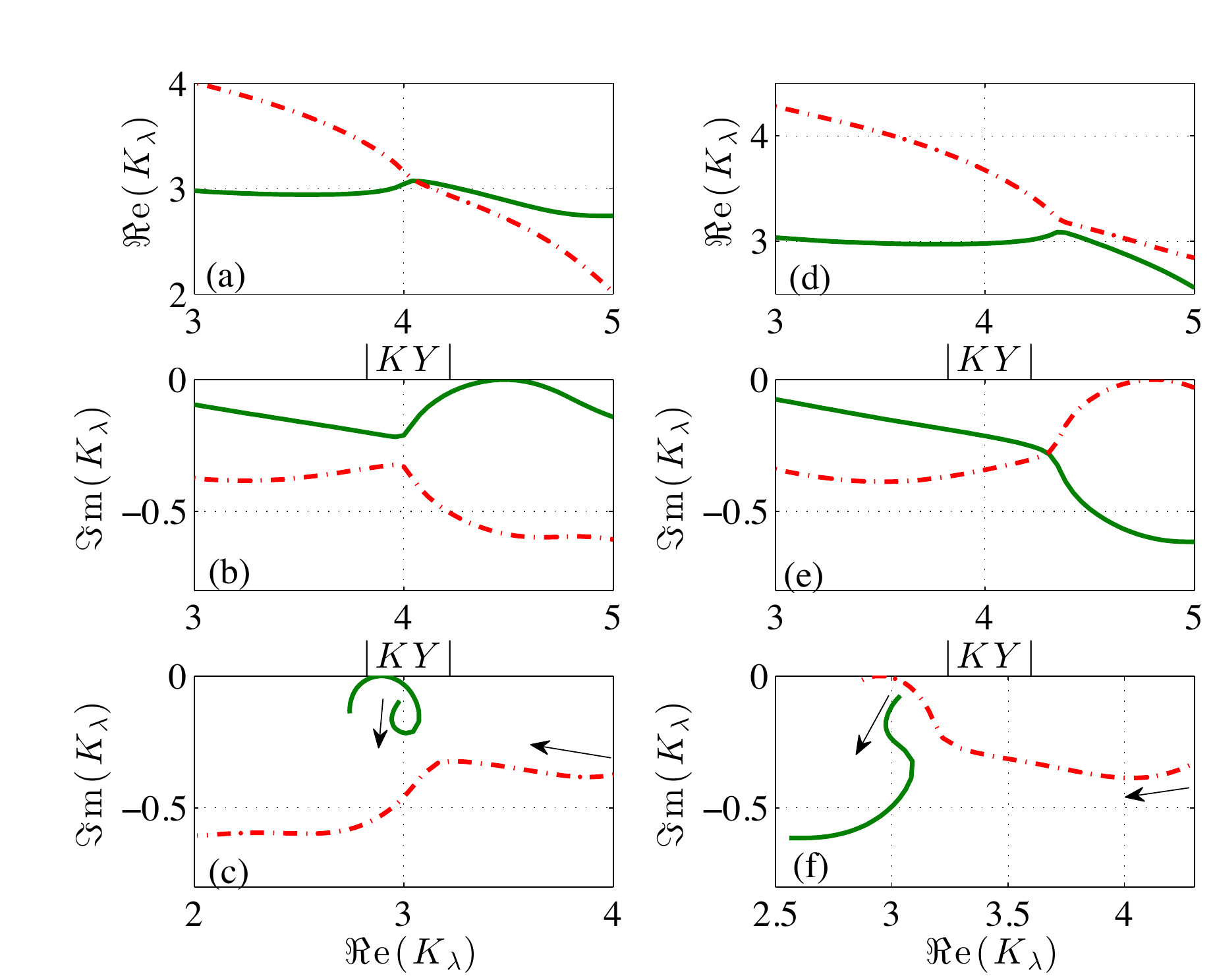}
 \caption{(Color online) Two types of  avoided crossings of $K_{\lambda}$  as a function $\vert KY\vert$ with  $K=2.5$ under different values of $a/h$: (a), (b) and (c) for $a/h=3.8$; (d), (e) and (f) for $a/h=3.6$. They have the same two modes as avoided crossing ``A'' in Fig.\;\ref{Figure3}. EP occurs at $\vert KY\vert=4.18$ and $a/h=3.7$, labelled by ``$C_1$" in Fig.\;\ref{Figure5}.  At  (b) $\vert KY\vert=4.5$ and (e) $\vert KY\vert=4.8$, $\Im $m$(K_\lambda$) goes to zero. $K_\lambda$ turns to be a real resonance frequency.  }
\label{Figure4}
\end{figure}

As we mentioned before, the matrix $\mathsf{H_{eff}}$ depends on three parameters.  Another parameter, the liner length $a/h$, is used here to manipulate the avoided crossing. Taking  the same two modes with avoided crossing ``A" in Fig.\;\ref{Figure3} as an example, we decrease the liner length $a/h$, the results are shown in Fig.\;\ref{Figure4}.  When $a/h$ varies, there is a jump between two types of avoided crossings.  When $a/h=3.8$, there is a crossing for $\Re$e($K_\lambda$) and not for $\Im$m($K_\lambda$) (see Fig. \ref{Figure4} (a) and (b)), while for $a/h=3.6$, there is no crossing for $\Re$e($K_\lambda$) and crossing for $\Im$m($K_\lambda$) (see Fig. \ref{Figure4} (d) and (e)).  

 \medskip
 
These findings and the type change of the avoided crossings from Fig. \ref{Figure4} (c) to (f) suggest that there should  exist a critical value $a_{cri}/h$, with which the curves of $K_\lambda$ as a function of $\vert KY\vert$ will cross at a critical value $\vert KY\vert_{cri}$. We found that at $\vert KY\vert=4.18$, $a/h=3.7$, and $K=2.5$, the two eigenvalues coalesce, so do  their eigenfunctions. ($a_{cri}/h,\; \vert KY\vert_{cri}$) is called an Exceptional Point (EP) in the parameter plane ($a/h,\; \vert KY\vert$). A strong mixing of the eigenfunctions of the two  modes occurs near the EP. The two modes which participate in the avoided crossing exchange their identities\cite{leissa, kuttler}.

 \medskip
 
\begin{figure}[h!]
\centering
\includegraphics[height=6.5cm,width=8cm]{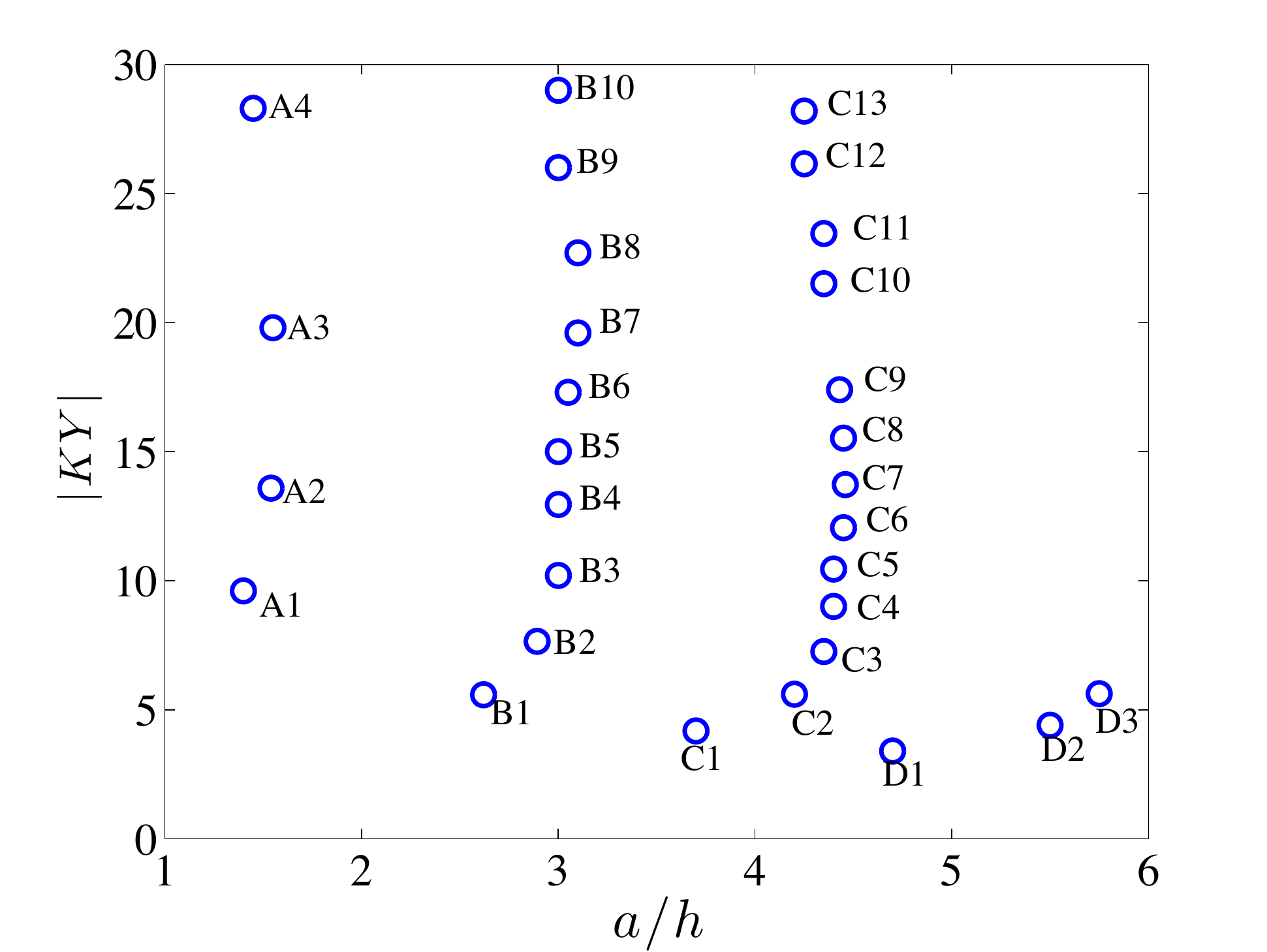}
\caption{(Color online) The distribution of EPs of $\mathsf{H_{eff}}$ in the ($a/h$, $\vert KY\vert $) plane with $K=2.5$. The branches A and C (B and D) correspond to the crossings between two modes in closed cavity with even (odd) symmetry about $x$.}
\label{Figure5}
\end{figure}

Avoided crossing occurs in the vicinity of an EP.  EPs were first introduced by Kato\cite{kato}, and were extensively developed by Heiss\cite{heiss_avoided_1990, heiss_repulsion_2000, heiss_avoided_2004}, Rotter\cite{rotter_non-hermitian_2009, rotter1, rotter2}, and Berry\cite{berry}. The mathematically topological structures of Riemann sheets at an EP, depend on a complex or two real parameters. A typical EP distribution of $\mathsf{H_{eff}}$ in the plane of $\vert KY\vert$ and $a/h$, with $K$  fixed, is shown in Fig.\;\ref{Figure5}. They occur near the crossings between two modes in the closed cavity that have the same symmetry about $x$. The branches A and C (B and D) correspond to the crossings between two modes in the closed cavity with even (odd) symmetry about $x$.

 \medskip

It is remarkable that at $\vert KY\vert=4.5$ in Fig.\;\ref{Figure4} (b) and $\vert KY\vert=4.8$ in Fig.\;\ref{Figure4} (e), $\Im$m($K_\lambda$) goes to zero. This is a real resonance frequency in the \textit{open} scattering region \Rmnum{2}. It is this kind of mode that Friedrich \textit{et al}\cite{friedrich_interfering_1985} found in nuclear reaction, and Sadreev \textit{et al.}\cite{sadreev_bound_2006} found in quantum billiards. This is the trapped mode that plays a crucial role in a transmission zero in the lined waveguide.  

 \medskip

\begin{figure}[h!]
\centering
\includegraphics[height=7.5cm,width=8.3cm]{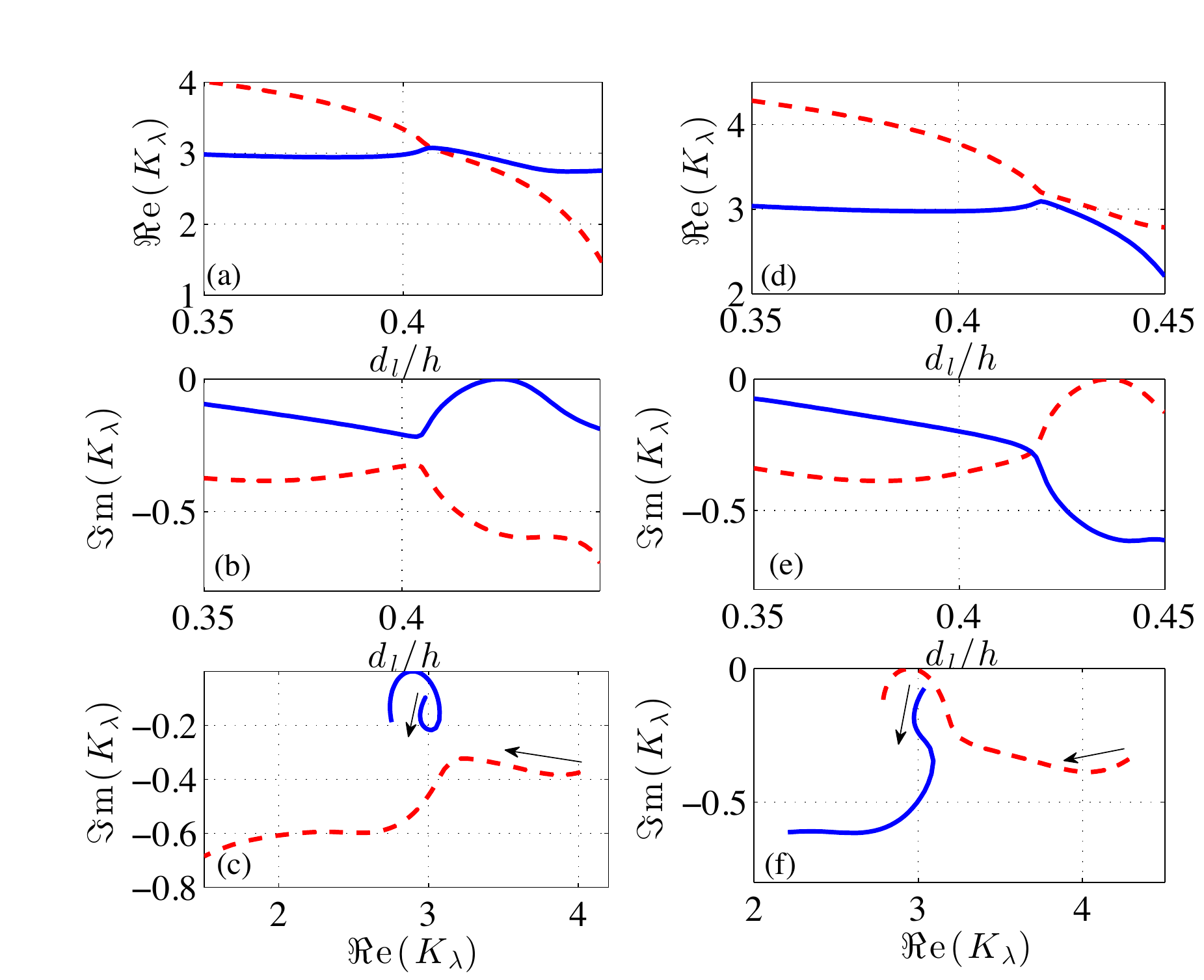}
\caption{(Color online)  Re-plot Fig.\;\ref{Figure4} using $d_l/h$, the liner depth, as x-coordinate.}
\label{Figure6}
\end{figure}

To realize a practical design of transmission zero, we need to understand the modes behaviour in the parameters space ($a/h$, $d_l/h$, $K$). We re-produce Fig.\;\ref{Figure4} in Fig.\;\ref{Figure6} using $d_l/h$ as a varying parameter. It is quite surprising  that the eigenvalue trajectories in the two  different parameter spaces of Figs.\;\ref{Figure4}\;and\;\ref{Figure6} are very similar to each other. We find that at $a/h=3.8$ and $d_l/h=0.425$, real resonance (trapped mode) occurs, with $K=2.5$.  

 \medskip
 
\begin{figure}[h!]
\centering
\includegraphics[height=7cm,width=8.3cm]{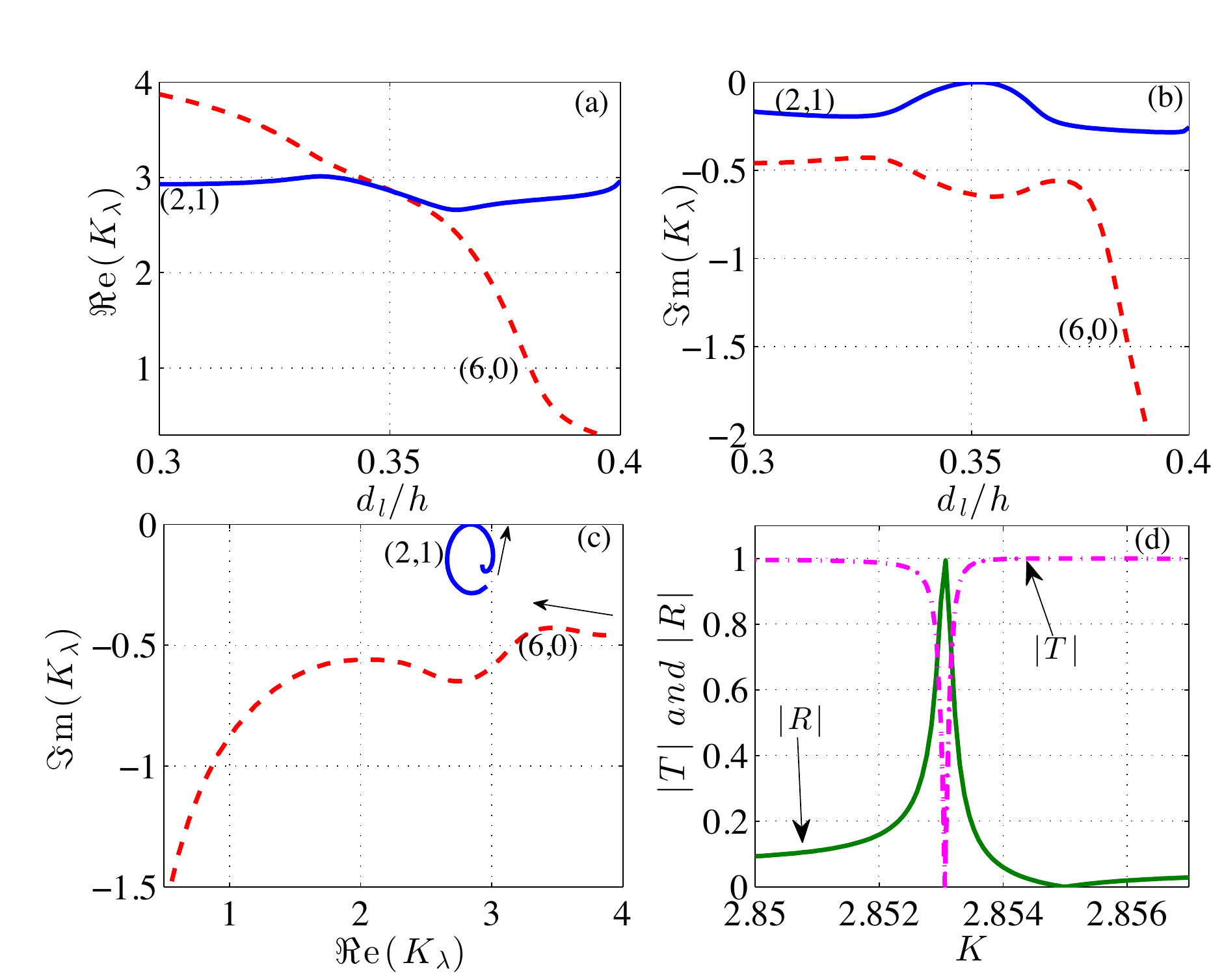}
\caption{(Color online) Transmission and reflection coefficients of the plane mode and trajectories of eigenvalues $K_\lambda$ of matrix $\mathsf{H_{eff}}$ as a function of $d_l/h$ with $K=2.85, a/h=3.8$, and $Re=0$: (a) real parts, (b) imaginary parts, (c) in complex $K_\lambda$ plane, and (d) transmission and reflection coefficients as a function of $K$ with $d_l/h=0.35, a/h=3.8$, and $Re=0$. In the figure, we also label the indices of the corresponding mode in closed cavity.} 
\label{Figure7}
\end{figure}

\begin{figure}[h!]
\centering
\includegraphics[height=3cm,width=8.3cm]{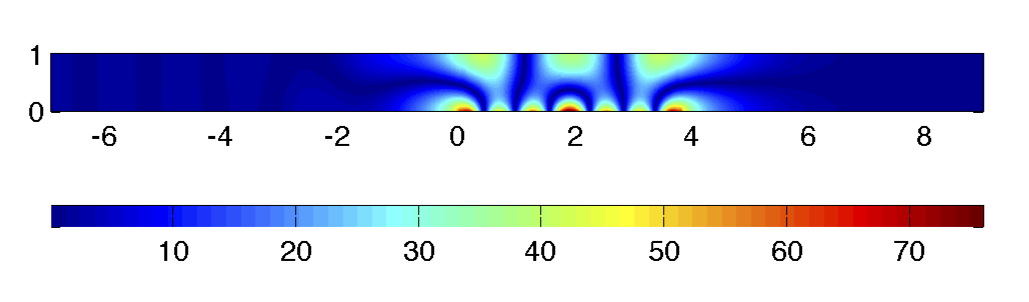}
\caption{(Color online)  The absolute value of sound pressure field in the waveguide when real resonance frequency is crossed.}
\label{Figure8}
\end{figure}

As pointed out earlier, $\mathsf{H_{eff}}$ depends on frequency $K$, so do $K_\lambda$. To find the frequency at which transmission zero occurs, we need to solve equation $K=K_\lambda(K)$. Therefore, near the EP labelled by ``$C_1$" in Fig.\;\ref{Figure5}, the real resonance frequency we seek for corresponds to the point where the curves $y=K_\lambda(K, d_l/h, a/h=3.8)$ cross the line $y=K$. We find that at $K=2.853$, $d_l/h=0.35$, and $a/h=3.8$, real resonance (trapped mode) occurs. Using this group of parameters, we calculate the transmission and reflection coefficients of the plane mode by Eq.\;(\ref{Eq_20}), as shown in Fig.\;\ref{Figure7}. A transmission zero occurs at $K=2.853$. The corresponding sound pressure field is shown in Fig.\;\ref{Figure8}. At the same time, reflection coefficient has a peak with amplitude $1$. 

 \medskip

\begin{figure}[h!]
\centering
\includegraphics[height=6cm,width=8.3cm]{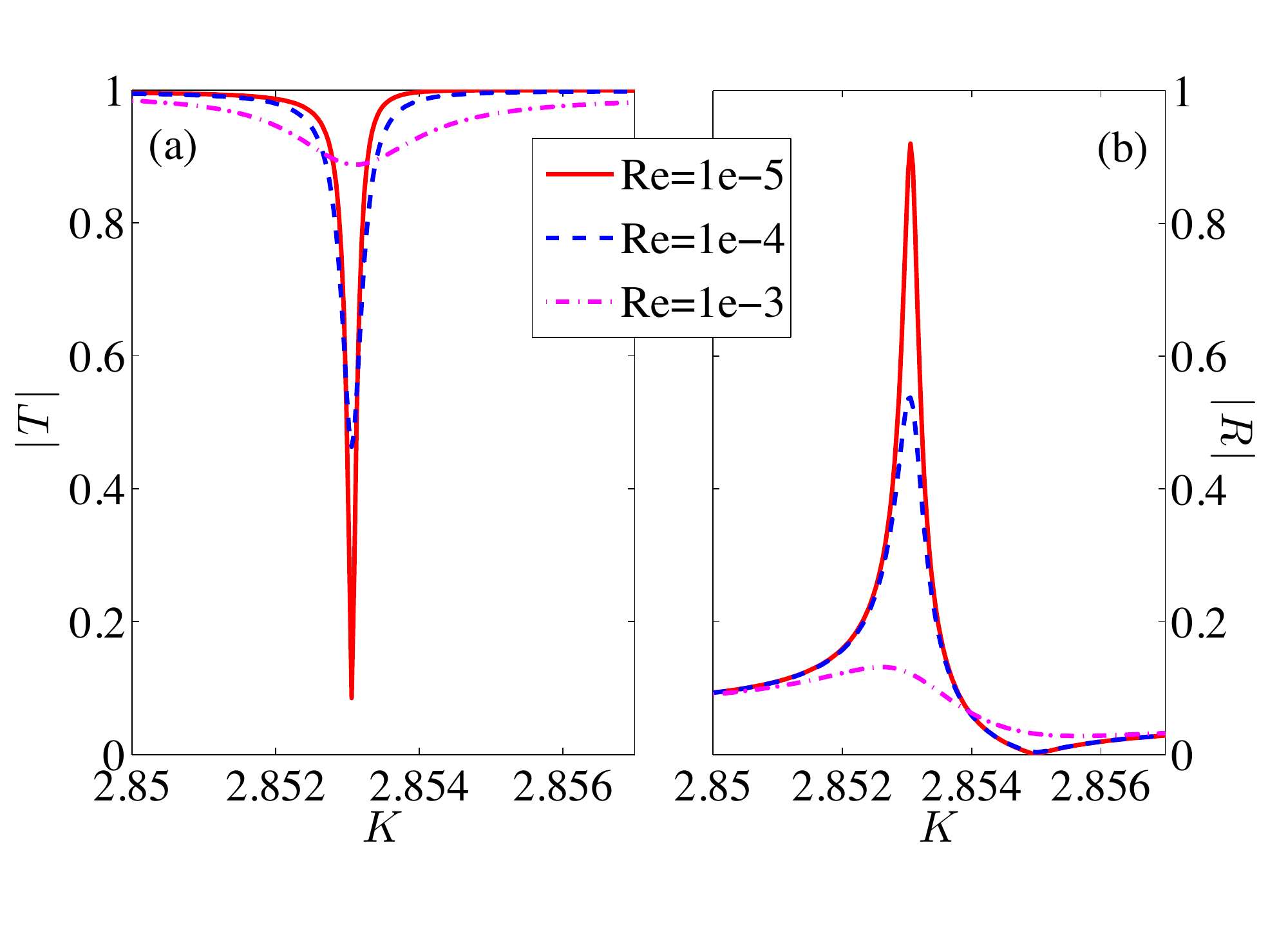}
\caption{(Color online) (a) Transmission and (b) reflection coefficients of the plane mode as a function of frequency under different values of resistance $Re$ with $d_l/h=0.35$. }
\label{Figure9}
\end{figure}

Now we consider the effects of dissipation. A small $Re$ is added in the impedance model Eq.\;(\ref{impedancemodel}). In Fig.\;\ref{Figure9}, we plot the reflection and transmission coefficients as a function of $K$ with $d_l/h=0.35$ and $a/h=3.8$ under different values of $Re$. The amplitude peaks of the reflection and transmission coefficients decrease rapidly with increasing $Re$. The high sensitivity on the dissipation is due to the fact that the crucial ingredient to form Fano resonance in the 
transmission and reflection coefficients -- the trapped mode with real resonance frequency,  turns to be quasi-trapped mode with complex resonance frequency. The sharp asymmetric profile becomes smoother, as shown in Fig.\;\ref{Figure9}.

\setlength{\parindent}{5ex}

\section{\label{sec4}CONCLUSIONS}
 
We have shown that the acoustic scattering matrix can be efficiently described with the help of an effective matrix $\mathsf{H_{eff}}$. This effective matrix is linked to the resonances of the scattering region opened to the infinite waveguides and gives the poles of the scattering matrix. Those poles are in the vicinity of exceptional points.

 \medskip
 
There are an infinite number of exceptional points in the parameters plane ($a/h$, $\vert KY\vert$), at which the eigenvalues and eigenfunctions of two modes coalesce, for the \textit{open} lined section. In the vicinity of each exceptional point, crossing or avoided crossing of the real and imaginary parts of the complex resonances (eigenvalues) occur. By varying one of the parameters, say $\vert KY\vert$, one mode turns to be a trapped mode, its resonance frequency becomes real. When a plane mode is incident, a transmission zero is present in the vicinity of a resonance peak, called Fano resonance, in the transmission amplitudes when the real resonance frequency is crossed. 

 \medskip
 
EPs, real resonance frequencies, and transmission zeros can also be obtained for parameters ($a/h$, $d_l/h$), which can suit the practical need of noise mitigation.  We have also shown that the transmission zeros and peaks are highly sensitive to the dissipation.   

 \medskip
 
With the aid of the eigenvalues and eigenfunctions of matrix $\mathsf{H_{eff}}$, the traditional acoustic resonance scattering formula is extended to describe the coupling effects between the open lined section and the rigid parts of the waveguide.

\medskip
In this paper, the numerical calculations are made for one incident mode and thus ``zero transmission" means that the total sound field is stopped. However, the model is valid for multimode being incident. To stop the total sound field for multimode propagation is still an open question.

 \noindent \textbf{Acknowledgements}
 
 \setlength{\parindent}{0.7cm} 
We gratefully acknowledge the support from the European Union through ITN-project FlowAirS (Contract No. FP7-PEOPLE-2011-ITN-289352).

\appendix
\section*{\label{Appen_A} Appendix: Derivation of Eq.\;(\ref{bound-general}) and (\ref{S4})}
\subsection*{1. Derivation of Eq. (\ref{bound-general})}
Multiplying Eq.\;(\ref{goven}) by $\bm\psi$,  integrating over the closed cavity, we obtain
\begin{equation}
\begin{aligned}
\int_0^1\int_0^{a/h}\bm\psi\left(\frac{\partial^2 p}{\partial x^2}+\frac{\partial^2 p}{\partial y^2}\right)dxdy=-K^2\int_0^1\int_0^{a/h} \bm\psi p \;dxdy.
\end{aligned}
\label{Eq_Ap2}
\end{equation}
Applying Green's theorem for Eq.\;(\ref{Eq_Ap2}), substituting  Eq.\;(\ref{rigid cavity}) into the resulting equation, we have
\begin{equation}
\begin{aligned}
&\int_0^1\left.\left( \bm\psi\frac{\partial p}{\partial x}-\frac{\partial \bm\psi}{\partial x}p\right)\right\vert_{x=0}^{x=a/h}dy+\int_0^{a/h}\left.\left( \bm\psi\frac{\partial p}{\partial y}-\frac{\partial \bm\psi}{\partial y}p\right)\right\vert_{y=0}^{y=1}dx\\
&=-(K^2\mathsf{I}-\mathsf{\Gamma)}\int_0^1\int_0^{a/h} \bm\psi p\; dxdy,
\end{aligned}
\label{Eq_Ap3}
\end{equation}
where $\mathsf{I}$ is an identity matrix, and $\mathsf{\Gamma}$ is a diagonal matrix with elements $\gamma_{\mu\nu}^2$.
Substituting the boundary conditions, Eqs.\;(\ref{BC_III}) and (\ref{rigid cavity bound}),  into Eq.\;(\ref{Eq_Ap3}), results in
\begin{equation}
\int_0^1\left.\left( \bm\psi\frac{\partial p}{\partial x}\right)\right\vert_{x=a/h}^{x=0}dy-\text{j}KY\int_0^{a/h} \bm\psi(x,0) p(x,0)dx=(K^2\mathsf{I}-\mathsf{\Gamma)}\int_0^1\int_0^{a/h} \bm\psi p\; dxdy.
\label{Eq_A5}
\end{equation}
Replacing the pressure function $p$ inside the scattering region by Eq.\;(\ref{p_in}), and using the orthogonality property of eigenfunctions $\psi_{\mu\nu}$,  the expansion coefficients $\bm a$ can be written as
\begin{equation}
\bm a =\left[K^2\mathsf{I}-\mathsf{\Gamma}+\text{j}KY\int_0^{a/h} \bm\psi(x,0) \bm\psi^T(x,0)dx\right]^{-1}\int_0^1\left.\left( \bm\psi\frac{\partial p}{\partial x}\right)\right\vert_{x=a/h}^{x=0}dy.
\label{Eq_Ap5}
\end{equation}
Substitute the upper expression of $\bm{a}$ into Eq.\;(\ref{p_in}),  we end up with Eq.\;(\ref{bound-general}).

\subsection*{2. Derivation of Eq. (\ref{S4})}
Due to  the symmetry property  of the matrix $\mathsf{S}$, Eq.\;(\ref{Eq_20}) can be rewritten as
\begin{equation}
\mathsf S=-\mathsf {I_{2M}}+\frac{2\text{j}\mathsf{C_{0a}}^T(K^2\mathsf{I}-\mathsf{H_{in}})^{-1}\mathsf{C_{0a}}\mathsf{K^x_{2M}}}{\mathsf{I_{2M}}+\text{j}\mathsf{C_{0a}}^T(K^2\mathsf{I}-\mathsf{H_{in}})^{-1}\mathsf{C_{0a}}\mathsf{K^x_{2M}}}.
\label{Eq_A1}
\end{equation}
Now we expand the denominator in Eq. (\ref{Eq_A1}) into a geometric series \cite{stockmann_effective_2002}, 
\begin{equation}
\begin{aligned}
\mathsf{S}&=-\mathsf{I_{2M}}+2\text{j}\mathsf{C_{0a}^T}\frac{1}{K^2\mathsf{I}-\mathsf{H_{in}}}\mathsf{C_{0a}K_{2M}^x}\sum_{q=0}^{\infty}\left(-\text{j}\mathsf{C_{0a}^T}\frac{1}{K^2\mathsf{I}-\mathsf{H_{in}}}\mathsf{C_{0a}K_{2M}^x}\right)^q\\
&=-\mathsf{I_{2M}}+2\text{j}\mathsf{C_{0a}^T}\frac{1}{K^2\mathsf{I}-\mathsf{H_{in}}}\sum_{q=0}^{\infty}\left(-\text{j}\mathsf{C_{0a}K_{2M}^x}\mathsf{C_{0a}^T}\frac{1}{K^2\mathsf{I}-\mathsf{H_{in}}}\right)^q\mathsf{C_{0a}K_{2M}^x}\\
&=-\mathsf{I_{2M}}+2\text{j}\mathsf{C_{0a}^T}\frac{1}{K^2\mathsf{I}-\mathsf{H_{in}}}\frac{1}{1+\text{j}\mathsf{C_{0a}K_{2M}^x}\mathsf{C_{0a}^T}\frac{1}{K^2\mathsf{I}-\mathsf{H_{in}}}}\mathsf{C_{0a}K_{2M}^x}\\
&=-\mathsf{I_{2M}}+2\text{j}\mathsf{C_{0a}^T}\frac{1}{K^2\mathsf{I}-\mathsf{H_{in}}+\text{j}\mathsf{C_{0a}K_{2M}^x}\mathsf{C_{0a}^T}}\mathsf{C_{0a}K_{2M}^x}.
\end{aligned}
\label{Eq_A}
\end{equation}
By the expression of $\mathsf{K_N}$, the upper equation results in the Eq. (\ref{S4}).

\newpage
\renewcommand*\numberline[1]{FIG.\,#1\space}

 \renewcommand*\listfigurename{LIST  OF  FIGURES}   
 \lhead{} 
 
\listoffigures

\end{document}